\documentclass{emulateapj}

\usepackage{natbib}

\newcommand{\gs}{\rm \hspace{0.3em}\raisebox{0.4ex}{$>$}\hspace{-0.75em}\raisebox{-.7ex}{$\sim$}\hspace{0.3em}}

\slugcomment{}

\shorttitle{The core-cusp problem in cold dark matter halos}
\shortauthors{Ogiya \& Mori}

\begin{document}


\title{The core-cusp problem in cold dark matter halos and supernova feedback: Effects of Mass Loss}


\author{Go Ogiya and Masao Mori}
\affil{Graduate School of Pure and Applied Science, University of Tsukuba, 1-1-1 Tennodai, Tsukuba 305-8577, Japan}
\email{ogiya@ccs.tsukuba.ac.jp}

%
%
%
%

\begin{abstract}
The core-cusp problem remains as one of the unsolved discrepancies between observations and theories predicted by the standard paradigm of cold dark matter (CDM) cosmology. 
To solve this problem, we perform $N$-body simulations to study the nonlinear response of CDM halos to the variance of the gravitational potential induced by gas removal from galaxy centers. 
In this study, we focus on the timescale of the gas ejection, which is strongly correlated with stellar activities, and demonstrate that it is one of the key factors in determining the dynamical response of CDM halos. 
The results of simulations show that the power-low index of the mass-density profile of the dark matter halo correlated with the timescale of the mass loss, and it is flatter when the mass loss occurs over a short time than when it occurs over a long time. 
However, it is still larger than typical observational values; in other words, the central cusp remains for any mass loss model in the simulations. 
Moreover, for the slow mass-loss case, the final density profile of the dark matter halo recovers the universal density profiles predicted by the CDM cosmology.
Therefore, mass loss driven by stellar feedback may not be an effective mechanism to flatten the central cusp.
\end{abstract}





%
%
%
%

\section{Introduction}
Cold dark matter (CDM) cosmology is the standard paradigm of structure formation in the universe. However, it has several serious-unsolved problems.
Recent observations of nearby dwarf galaxies and low surface brightness galaxies (LSBs) have revealed that the density profile of the DM halo is constant at the center of such galaxies (e.g., Moore 1994; Burkert 1995; de Blok et al. 2001; Swaters et al. 2003; Spekkens et al. 2005).
In contrast, cosmological $N$-body simulations based on collisionless CDM have always predicted a steep power-law mass-density distribution at the center of CDM halos (e.g., Navarro et al. 1997; Fukushige \& Makino 1997; Moore et al. 1998).
This discrepancy is the well-known core-cusp problem of the CDM scenario.
To solve this problem, so far, numerical simulations and analytical methods used to study the dynamical response of DM halos to the variance of the gravitational potential.
Supernova feedback to the interstellar medium plays a significant role on forming less massive galaxies, because such galaxies have a shallower gravitational potential than giant galaxies.
The effects of stellar feedback differ substantially from one galaxy to another depending on the gravitational potential (Mori et al. 1999; MacLow \& Ferrara 1999).
In low-mass dwarf galaxies, supernovae blow gas out from the galaxy centers.
This mass loss makes the gravitational potential around the center of the DM halos shallower, and the mass-density distribution has a flat core (Navarro et al. 1996a; Read \& Gilmore 2005).
These previous works have always assumed that mass loss from the center of DM halos occurs instantaneously or within a fixed period.
However, the timescale of mass loss may be also a key factor in determining the dynamical response of DM halos.
Furthermore, each dwarf galaxy in the local group has a unique star formation history (Mateo 1998; Weisz et al. 2011).
De Souza et al. (2011) analytically demonstrated that the energy gain of particles flattens the central cusp; however, they assumed that each particle receives an equal amount of energy.
This assumption must be verified by numerical simulations with sufficient accuracy.
Therefore, in this paper, we run collisionless $N$-body simulations that focus on the dependence of the dynamical response of DM halos on the timescale of gravitational potential variance.
In other words, we study how the dynamics of DM halos depend on star formation activities.
We demonstrate that the timescale of mass loss is one of the important factors in determining the dynamical response of DM halos.
Finally, we discuss the surprising result: the mass loss of the baryon is not an effective mechanism for flattening the central cusp.

The structure of this paper is as follows. In \S 2, we describe the numerical simulations. In \S 3, we show the simulation results. Finally, we discuss the results in \S 4.

%
%
%
%

\section{Numerical Models}


\subsection{The DM halo model}

The density distribution of a DM halo obtained from $N$-body simulations based on CDM cosmology is well fitted by the following:
\begin{equation}
\rho_{\rm DM}(r) = \frac{\rho_0 R_{\rm DM}^3}{r^{\alpha}(r+R_{\rm DM})^{3-\alpha}} \label{DM_halo}
\end{equation}
where $r$ is the distance from the center of a DM halo, $\alpha$ is a power-low index, and $\rho_0$ and $R_{\rm DM}$ are the scale density and scale length of the DM halo, respectively. The density distribution changes from $\rho \propto r^{-\alpha}$ in the center of a DM halo ($r < R_{\rm DM}$) to $\rho \propto r^{-3}$ its outskirts ($r > R_{\rm DM}$). Here, $\alpha = 1.0$ corresponds to the Navarro--Frenk--White (NFW) model (Navarro et al. 1996b; Navarro et al. 1997) and $\alpha = 1.5$ corresponds to the Fukushige--Makino--Moore (FMM) model (Fukushige \& Makino 1997; Moore et al. 1999).
To generate equilibrium $N$-body systems, we use the fitting formulation of the distribution function (DF) proposed by Widrow (2000). In this case, the DF only depends on energy, and the velocity dispersion of the system is isotropic.


\subsection{The baryon model}

Hernquist model (Hernquist 1990) is frequently used to describe bulges or elliptical galaxies, because its surface brightness profile matches de Vaucouleurs law (de Vaucouleurs 1948). Therefore, we adopt the Hernquist potential to describe the baryon potential around the center of DM halos. The external potential is given by
\begin{equation}
\Phi_{\rm b} (r) = -\frac{G M_{\rm b}}{r+R_{\rm b}} \label{Hernquist}
\end{equation}
where $G$ is the gravitational constant, and $M_{\rm b}$ and $R_{\rm b}$ are the baryon mass and scale length of the external potential, respectively. 
To generate the initial condition, we relax the equilibrium $N$-body system quoted above in the external baryon potential for $\sim 30 ~t_{\rm d}$, where $t_{\rm d}(r)$ is the dynamical time defined by
\begin{equation}
t_{\rm d}(r) = \sqrt{\frac{3 \pi}{32 G \bar{\rho} (<r_{\rm d})}} \label{dynamical_time}
\end{equation} 
where $ \bar{\rho} (<r) $ is the average density of a DM halo within a radius $r=0.2$ kpc throughout this paper.
Then, to represent baryon ejection from the center of galaxies, we change the baryon mass $M_{\rm b}=M_{\rm b, tot}(1-t/T_{\rm out})$, where $M_{\rm b,tot}$ and $T_{\rm out}$ are the total baryon mass and the timescale of mass loss. 
For the instantaneous mass-loss model, we simply set $M_{\rm b}=0$.


\subsection{Description of runs}

In this paper, we simulate the dynamical response of a DM halo with the virial mass $M_{\rm vir} = 10^9 M_{\odot}$, the virial radius $R_{\rm vir} = 10$ kpc, and the scale length $R_{\rm DM} = 2$ kpc. In this case, the mean dynamical time within 200 pc is $t_{\rm d} = 4$ Myr for the FMM model and 10 Myr for the NFW model. The total baryon mass is $M_{\rm b,tot} = 1.7 \times 10^8 M_{\odot}$, and the scale length of the baryon potential is $R_{\rm b} = 0.04$ kpc. Here, we assume that the total baryon mass included in the dwarf galaxies is initially 17\% of the halo mass. This fraction matches the cosmic baryon fraction $f_{\rm b}$ obtained from Wilkinson Microwave Anisotropy Probe (WMAP) observations (Spergel et al. 2007; Komatsu et al. 2009). The adopted parameters for each of our simulations are given in Table 1.
Throughout this paper, the opening parameter of the Barnes--Hut tree algorithm (Barnes \& Hut 1986) is $\theta = 0.8$ and the softening parameter is $\epsilon = 0.008$ kpc.

\begin{table}
\begin{center}
\caption{Summary of simulation run}
\begin{tabular}{cccccc}
ID & DM halo & $N$  & $T_{\rm rel}$ & Mass-loss ($T_{\rm out}$) & Fitted $\alpha$ \\
\tableline\tableline
UP1 & FMM & 1,048,576 & 495 $t_{\rm d}$ & -             &  \\ 
UP2 & FMM & 16,384    & 7.74 $t_{\rm d}$ & -             &  \\ 
UP3 & NFW & 1,048,576 & 148 $t_{\rm d}$ & -             &  \\ 
UP4 & NFW & 16,384    & 2.32 $t_{\rm d}$ & -             &  \\  
ML1 & FMM & 1,048,576 & 495 $t_{\rm d}$ & instantaneous & 0.85 \\
ML2 & FMM & 1,048,576 & 495 $t_{\rm d}$ &   1 $t_{\rm d}$ & 1.20 \\ 
ML3 & FMM & 1,048,576 & 495 $t_{\rm d}$ &  10 $t_{\rm d}$ & 1.44 \\
ML4 & FMM & 1,048,576 & 495 $t_{\rm d}$ &  50 $t_{\rm d}$ & 1.46 \\ 
ML5 & NFW & 1,048,576 & 148 $t_{\rm d}$ & instantaneous & 0.42 \\ 
ML6 & NFW & 1,048,576 & 148 $t_{\rm d}$ & 50 $t_{\rm d}$  & 0.89 \\ 
\tableline
\end{tabular}
\tablecomments{ML indicates mass-loss runs, and UP indicates unperturbed runs to examine the stability of the $N$-body system and the effects of two-body relaxation. $T_{\rm rel}$ is a two-body relaxation time within 0.2 kpc.}
\end{center}
\label{table_mass-loss}
\end{table}

%
%
%
%

\section{Simulation results}

In this section, we present the simulation results showing the evolutionary processes in the instantaneous and quasi-adiabatic mass-loss models. 


\subsection{Dynamical evolution of DM halos}

We show how the density profile of a DM halo evolves after mass loss for the instantaneous mass-loss case with the FMM initial condition (ML1) at the left panels in Fig. 1. During the first $\sim 30 ~t_{\rm d}$ (see Fig. 1b), the system expands and the central cusp is flattened to the core. In this phase, the bulk of particles in the center of DM halo moves outward, and then the self-gravity of the DM halo slows the particles.
After 30 $t_{\rm d}$, the particles lose their radial velocity and fall back into the center. In Fig. 1c-1d, these particles assemble at the center of the halo, and the cusp is regenerated. This cusp regenerating process occurs from the center to the outward, and finally, the system reaches the quasi-equilibrium state (see Fig 1e). We emphasize that the central cusp has been regenerated, but the power-law index of density at the center $\alpha = 0.85$ is smaller than the initial condition $\alpha = 1.5$. A similar phenomenon occurs for the instantaneous mass-loss case with the NFW initial condition (ML5): the density at the center is $\alpha = 0.42$ for the initial condition $\alpha = 1.0$. Therefore, instantaneous mass-loss flattens the central cusp in the density distribution to some degree.

In the right panels of Fig. 1, we show how the density profile of a DM halo evolves during and after mass loss for the slow (quasi-adiabatic) mass-loss case with the FMM initial condition (ML4).
Before the mass-loss operation, the DM halo has a high central density because of the baryon potential; this density decreases with decreasing baryon mass. Slow mass-loss does not significantly flatten the central cusp in the density distribution.
Moreover, the system always keeps the virial state for the slow mass-loss process, and the final density profile of the DM halo recovers the FMM initial condition. A similar phenomenon occurs for the adiabatic mass-loss case with the NFW initial condition (ML6). 

\begin{figure}


\plotone{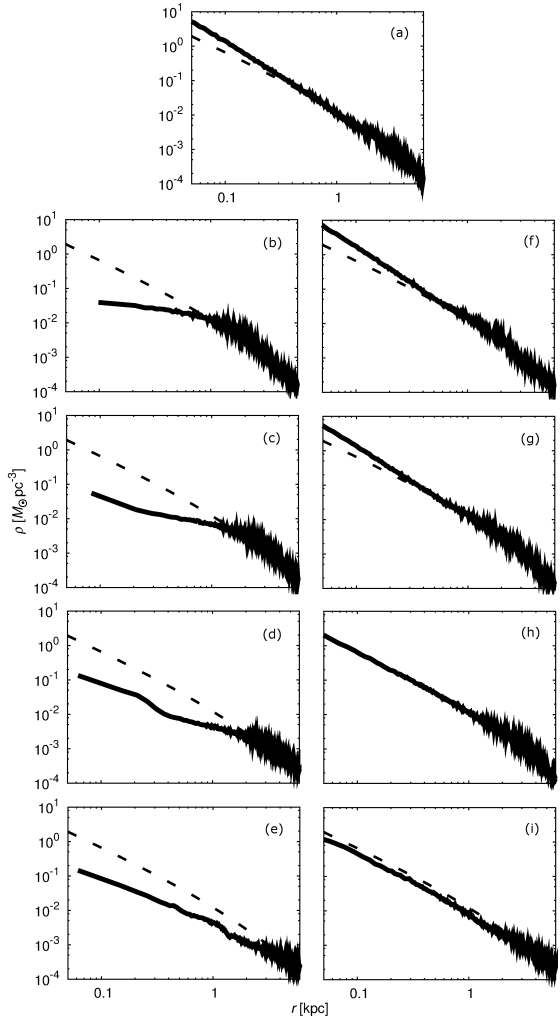}
\caption{
Evolution of the density profiles of a DM halo for the instantaneous mass-loss model (ML1; left panels) and the adiabatic mass-loss model (ML4; right panels). The top panel (a) shows the density profile of the quasi-equilibrium state after adding the external potential on the initial FMM model. The other panels show the density profiles of a DM halo at $15 ~t_{\rm d}$ (b and f), $30 ~t_{\rm d}$ (c and g), $50 ~t_{\rm d}$ (d and h) and $110 ~t_{\rm d}$ (e and i), respectively. Dashed-line represents the FMM initial condition.
}
\end{figure}


\subsection{Final states of DM halos}

Figure 2 shows the results of simulations for the FMM model (left panels) and the NFW model (right panels). The upper left (right) panel shows the density profile of a DM halo for the unperturbed model with the FMM (NFW) initial condition, respectively. The thick-solid line corresponds to the high-resolution run (UP1 and UP3) and the thin-solid line corresponds to the low-resolution run (UP2 and UP4).
In unperturbed runs with only a small number of particles, even though no external potential exists, a cusp-to-core transition caused by two-body relaxation occurs. Previous studies could not rule out this effect, because they did not use a sufficient number of particles in the central region. However, two of unperturbed runs clearly show that we have a sufficient number of particles to suppress the effect of two-body relaxation and the system remains stable at least several 100 dynamical times.

The middle panels in Fig. 2 show that the density profile of a DM halo has reached the quasi-equilibrium state after mass loss in ML runs. 
The thick-solid lines represent the instantaneous mass-loss model (ML1 and ML5) and the thin-solid line corresponds $T_{\rm out} = 50 t_{\rm d}$ (ML4 and ML6). 
We fit the quasi-equilibrium density profiles for all of ML runs using the least-squares method, and we find the following values for the parameter $\alpha$. For the FMM model, $\alpha$ is 0.85 for the instantaneous mass-loss (ML1), 1.20 for $T_{\rm out} =  t_{\rm d}$ (ML2), 1.44 for $T_{\rm out} = 10 ~t_{\rm d}$ (ML3), and 1.46 for $T_{\rm out} = 50 ~t_{\rm d}$ (ML4). In Table 1, we summarize the resultant $\alpha$ including for the NFW model. 
It is clear that mass loss occurs in a short timescale; in other words, intense stellar activities cause the density profiles of DM halos to become shallower. 
In the slow mass-loss cases (ML3, ML4 and ML6; $T_{\rm out} \gs 10 ~t_{\rm d}$), $\alpha$ is approximately conserved. Therefore, to flatten the central cusp needs intense stellar activity; however, even for cases ML1 and ML5 that have such intense stellar activity, $\alpha$ is still larger than the typical observational values of $\alpha \sim 0.2 - 0.3$ (Spekkens et al. 2005).

The lower panels of Fig. 2 show the amount of energy gained from the external baryon potential. While the baryon potential is added into DM halos, particles in the system acquire kinetic energy to balance the potential energy; conversely, during mass loss, particles lose kinetic energy. After mass-loss operation has finished, the total energy conserves. The lower panels of Fig. 1 demonstrate that after mass loss in the instantaneous mass-loss cases (ML1 and ML5; thick solid lines), DM halos gain more energy than that gained in the slower mass-loss cases (ML2 and ML6; thin solid lines). Therefore, the shorter the timescale of mass loss, the farther the DM halos expand and the flatter they become.

\begin{figure}


\plotone{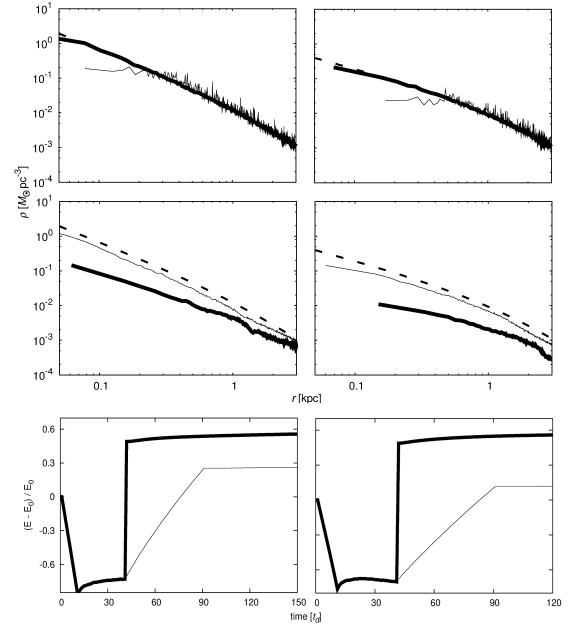}
\caption{
The upper left panel shows the density profile of a DM halo for the unperturbed (UP) model (solid lines) with the FMM initial condition (long-dashed line). The thick-solid line represents $N = 1,048,576$ (UP1) and the thin-solid line represents $N = 16,384$ (UP2).
The middle left (right) panel shows the quasi-equilibrium density profile of a DM halo for the FMM (NFW) model with the mass-loss. The thick-solid line represents the instantaneous mass-loss model (ML1 and ML5), and the thin-solid line corresponds $T_{\rm out} = 50 t_{\rm d}$ (ML4 and ML6). The lower left (right) panel shows the variance of the total energy of the system for the FMM model (NFW model). Thick-solid line corresponds the instantaneous mass-loss model, and thin-solid line corresponds $T_{\rm out}=50 ~t_{\rm d}$, where $E_{\rm 0}$ and $E$ are the total energies of DM halos at the initial and given times, respectively.
}
\end{figure}

%
%
%
%

\section{Discussion}
\begin{figure}


\plotone{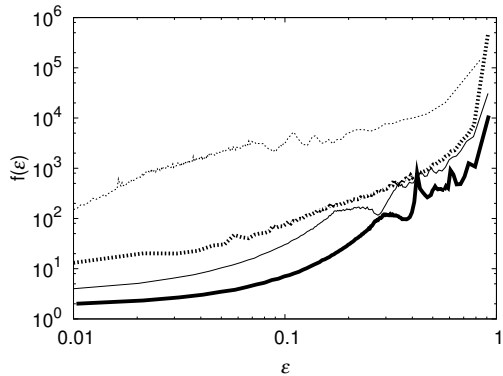}
\caption{
The distribution function of DM halos on the normalized particle energy. This energy is normalized by multiplying by the normalization factor $A$. Shown are the initial conditions for the FMM model (thick dashed line; $A=-0.70$), the quasi-equilibrium state before the mass-loss operation (thin dashed line; $A=-0.095$), the quasi-equilibrium state for the instantaneous mass-loss model ML1 (thick solid line; $A=-2.0$) and the quasi-adiabatic mass-loss model ML4 (thin solid line; $A=-1.05$).
}
\end{figure}

Previous numerical attempts to solve the core-cusp problem of the CDM scenario (Navarro et al. 1996a; Read \& Gilmore 2005) have been limited by two factors: (1) they did not consider the timescale of mass loss and (2) they could not remove the effects of the two-body relaxation, because they did not use the sufficient number of particles in the center of the DM halos. Our current numerical study overcomes these limitations by considering the timescale of mass loss and including a larger number of particles in the center of DM halos. We find that the timescale of mass loss is an important factor affecting DM halo dynamics.

Our study determines the dynamical response of DM halos to changes in the gravitational potential induced by stellar activities that remove the baryons. We especially focus on the timescale of gravitational variance. We find that the central cusp of the DM halo is flatter when mass loss occurs over a short time than when it occurs over a long time; this result is consistent with the analytical insight of Hills (1980). However, the power-law index of the central cusp $\alpha$ is still larger than typical observational values; in other words, the central cusp remains even for instantaneous mass-loss models. Therefore, mass loss may not be an effective mechanism to flatten the central cusp, at least in spherical systems. This suggests that DM halos may be stable to mass loss.

Figure 3 shows the DF for the FMM initial condition, the quasi-equilibrium state with the external baryon potential, ML1 and ML4. The shape of the DFs after mass loss is similar to the initial condition, suggesting that DM halos return to the initial condition not only in spatial-space but also in phase-space; therefore, DM halos are stable to mass loss. 

In this study, we find that a temporal cusp-to-core transition occurs for instantaneous mass-loss. 
After a few dozen dynamical times, the cusp regenerates from the center. 
However, this cusp has a smaller $\alpha$ than the initial conditions, suggesting that the cusp has been somewhat flattened by mass loss. 
In addition, we find that not only the density profile but also the energy distribution almost returns to the initial condition for adiabatic mass-loss.
The physical mechanism of this cusp regeneration has not been revealed, but it may aid in understanding the formation of the universal density profiles of CDM halos. 
Only in the linear perturbation regime, the Doremus--Feix--Baumann theorem (Doremus et al. 1971) states that collisionless $N$-body systems with DF $f(E)$, $df(E)/dE < 0$ are stable to radial perturbation. However, it is open question in the non-linear regime. 
In the series of subsequent studies, we will elucidate the detailled mechanism of the recovering cusp, and examine more realistic model including non-spherical analysis and gasdynamics.

\acknowledgments

We would like to thank M. Umemura, K. Yoshikawa, T. Okamoto, N. Kawakatsu, T. Kawaguchi, K. Hasegawa, and A. Tanikawa for useful discussions. Numerical simulations were performed with the FIRST simulator and T2K-Tsukuba at the Center for Computational Sciences at the University of Tsukuba. This work was supported in part by JSPS Grants-in-Aid for Scientific Research: (A) (21244013), (C) (18540242), and (S) (20224002), and Grants-in-Aid for Specially Promoted Research by MEXT (16002003).

\end{document}